\documentclass{ifacconf}
\usepackage{graphicx} 
\usepackage{natbib}

\usepackage{multirow}
\usepackage{color}
\usepackage{amsfonts}
\usepackage{amsfonts}
\usepackage{textcomp}
\usepackage{float}
\usepackage{graphicx}
\usepackage{bm}
\usepackage{amsmath}
\usepackage{amssymb}
\usepackage{url}
\usepackage{makecell}
\usepackage{xfrac}
\usepackage[ruled,vlined,algo2e]{algorithm2e}

\newcommand{\sgn}{\text{sgn}} 

\begin{document}
\begin{frontmatter}

\title{A New Simple-to-Configure Self-Perturbing Multivariable Extremum-Seeking Controller}

% Title, preferably not more than 10 words.

\thanks[footnoteinfo]{This research was supported wholly (or in part) by the U.S. Department of Defense, through the Environmental Security Technology Certification Program (ESTCP).}

\author[First]{Timothy I. Salsbury} 
\author[First]{Min Gyung Yu} 

\address[First]{Pacific Northwest National Laboratory, 902 Battelle Blvd, Richland WA 99354 USA (e-mail: \{timothy.salsbury; mingyung.yu\}@pnnl.gov)}

\begin{abstract}                % Abstract of 50--100 words
This paper presents a new stochastic {\em relay-based} extremum-seeking controller (ESC) for multi-input-single-output (MISO) systems. The goal of this work was to create an algorithm that is much simpler to configure than alternative approaches making deployment to real-world problems easier. A solution is developed first for a static map and then adapted for a general class of dynamic systems. The number of configurable parameters is one per input channel for the static case and only one additional parameter is needed for the dynamic version. The problem of gradient identification is solved via the use of stochastic relay gains and a simple stability proof for the static case is presented. Simulation tests demonstrate the performance of the strategy for optimizing both static and dynamic systems.
\end{abstract}

\begin{keyword}
extremum-seeking, adaptive control, real-time optimization, multivariable, stochastic
%Five to ten keywords, preferably chosen from the IFAC keyword list.
% real-time optimization \sep trim and respond \sep supervisory control \sep building energy use \sep HVAC systems
\end{keyword}

\end{frontmatter}
%===============================================================================

\section{Introduction}
Extremum-seeking control is a type of adaptive control for adjusting the parameters of a system in real-time to optimize a performance criterion. Single-input single-output (SISO) ESC is now a mature technology and there are several variants with the most popular being derived in the frequency domain and based on the use of external sinusoidal perturbation signals and the process of demodulation~(\cite{krstic2000stability}). 

Alternative approaches to extremum-seeking involve periodic switching~(\cite{korovin1974using}), which formalize early concepts of hill climbing as described by~\cite{blackman1962extremum}. These methods do not typically require an external perturbation and different sub-classes have been proposed based on either gradient-driven relays or sliding modes, e.g.,~\cite{olalla2007analysis,leyva2006mppt}.

Several other approaches based on switching have been proposed by~\cite{oliveira2011output, oliveira2012global, oliveira2007control}. In general, the utility of ESC is increased if it can be adapted to work for problems with multiple variables. Numerous papers have extended the SISO perturbation-based ESC to multiple variables, e.g.~\cite{ariyur2002analysis}, and many enhancements have been developed such as Newton-based search by ~\cite{ghaffari2012multivariable}. However, the application of periodic switching ESC has mostly been confined to SISO systems, although recent extensions to multivariable problems have been proposed by ~\cite{aminde2021multivariable, peixoto2020multivariable}.

ESC has been applied to many real-world problems as referenced by~\cite{tan2010extremum}; these include anti-lock braking systems, autonomous vehicles, energy systems, process control, and bioreactors among others. An important attribute of ESC is that it is model-free and therefore requires fewer parameters and engineering effort to configure compared with alternative approaches such as model-predictive control (MPC). However, configuration is still a barrier to the deployment of ESC for applications where expertise is low such as in buildings and energy efficiency applications~(\cite{dong2015self}).

For an SISO version of perturbation ESC, there can be up to six parameters required including integrator gain, low pass filter coefficient(s), high-pass filter coefficient(s), plus frequency, phase, and amplitude of the dither signal. In general, reported SISO relay-based ESC methods, such as in~(\cite{salsbury2020self}) require fewer parameters due to the elimination of the external perturbation signal. However, the previously cited multivariable switching ESC methods end up re-introducing periodic functions and include several additional configurable parameters as shown in~(\cite{aminde2021multivariable}). \cite{scheinker2024100} gives a recent comprehensive review of ESC methods.

In this paper, we develop a relay-based multivariable ESC with the primary objective of making it simple to configure. The presented algorithm builds on the SISO relay-ESC presented by~\cite{olalla2007analysis} and utilizes the multivariable chain rule with stochastic relay gains to ensure the gradient is identifiable. The resulting algorithm is applicable to MISO systems and has only one parameter per input channel and with one additional parameter needed for application to dynamic systems.

The paper is structured as follows: Section~\ref{sect:staticmap} develops the approach for a static map and includes a simple stability and convergence proof in~\ref{sect:stability}; Section~\ref{sect:dyanmic} extends the algorithm for application to dynamic systems; Section~\ref{sect:sim} presents results from simulation tests to demonstrate performance for both static and dynamic system cases; and Section~\ref{sect:conclusions} presents conclusions.

\section{Multi-Relay ESC for a Static Map}
\label{sect:staticmap}
\begin{figure}[ht]
\centering
\includegraphics[width=3in]{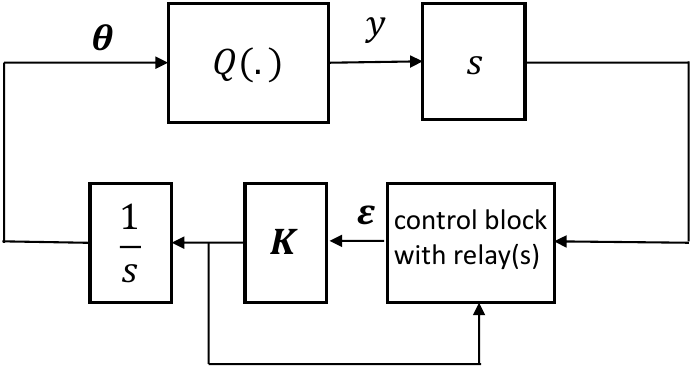}
\caption{Multi-relay ESC for static map}
\label{fig:ss_block}
\end{figure}

This paper is concerned with extending relay-based extremum-seeking control to multiple inputs. Figure~\ref{fig:ss_block} shows a block diagram of a viable strategy based on scaling up the SISO version to multiple channels.

In developing the approach, we will assume $p$ input channels and that the static map (cost function) is given by:
\begin{align}
y = Q(\boldsymbol{\theta}),~\boldsymbol{\theta} = [\theta_1,\theta_2, \ldots, \theta_p]^T,
\end{align}
with a local minimum at $\boldsymbol{\theta}^*$. The cost function is not known but can be measured via $y$ at any point in time and $\boldsymbol{\theta}$ can be manipulated. As a basis for extending the SISO relay approach to multiple inputs, we make use of the following multivariable chain rule:
\begin{align}
\label{eq:mchain}
    \frac{dy}{dt} = \frac{\partial y}{\partial \theta_1} \cdot \frac{d \theta_1}{dt} + \ldots \frac{\partial y}{\partial \theta_p} \cdot \frac{d \theta_p}{dt}.
\end{align}
Referring to Figure~\ref{fig:ss_block}, the time derivative of the cost function $\frac{dy}{dt}$ is the input to the control block and the outputs are the relay outputs $\boldsymbol{\epsilon}=[\epsilon_1,\ldots,\epsilon_p]^T$ where $\epsilon \in \{-1,1\}$ for all $\boldsymbol{\epsilon}$. The relay outputs are multiplied by a gain $\mathbf{K} = [k_1,\ldots,k_p]^T$ and then integrated and this integrated signal is the input to the static map. Note that the signal {\em before} the integrator block is thus equivalent to the time derivative of the input to the static map, i.e., $\frac{d \theta_1}{dt}|_{t=t_0} = \epsilon_{1,t_0} k_1$, etc. 

To move $\boldsymbol{\theta}$ toward the optimal point $\boldsymbol{\theta}^*$, we need to set the direction of the relays based on the partial derivatives $\frac{\partial y}{\partial \theta_1},\ldots,\frac{\partial y}{\partial \theta_p}$. However, inspection of Equation~\ref{eq:mchain} shows that these derivatives cannot be determined at one instant in time when there is more than one variable. A solution to this is to obtain estimates of all the time derivatives at different points in time. These different estimates can then populate rows in matrices as follows. First define $\mathbf{Y}^T = [\frac{dy(t)}{dt}|_{t=t_0},\ldots,\frac{dy}{dt}|_{t=t_0+i\Delta t}]$ as time derivatives of the cost function at times $t_0$ up to $t_0+i\Delta t$. Next define a matrix for the input channel time derivatives as:
$$ \mathbf{X} =
	\begin{bmatrix} 
	\frac{d\theta_1}{dt}|_{t=t_0} & \hdots &\frac{d\theta_p}{dt}|_{t=t_0} \\
	\vdots & \ddots & \vdots\\
	\frac{d\theta_1}{dt}|_{t=t_0+i\Delta t} & \hdots &\frac{d\theta_p}{dt}|_{t=t_0+i\Delta t} \\
	\end{bmatrix}
	\quad
	$$
and $\bf{g}^T=[\frac{\partial y}{\partial \theta_1},\hdots,\frac{\partial y}{\partial \theta_p}]$. Thus, the matrices are populated with time derivatives made at sequential points in time with a $\Delta t$ period of separation.

For $p$ input channels, a minimum of $p$ serial estimates of the time derivatives (corresponding to $i=p-1$) will need to be made to solve for $\mathbf{g}$. If this condition is satisfied, then the $\mathbf{g}$ vector can be obtained by solving the set of $p$ simultaneous equations to give:
\begin{align}
\label{eq:estimation}
    \hat{\mathbf{g}} = \mathbf{X}^{-1} \mathbf{Y}    
\end{align}

If we substitute the relay outputs and gains for the channel time derivatives the $\mathbf{X}$ can be re-stated as:
$$\mathbf{X} =
	\begin{bmatrix} 
	k_1 \epsilon_{1,t_0} & \hdots & k_p \epsilon_{p,t_0} \\
 	\vdots & \ddots & \vdots\\
	k_1 \epsilon_{1,t_0+i\Delta t} & \hdots & k_p \epsilon_{p,t_0+i\Delta t} \\
	\end{bmatrix}
	\quad
$$ where $\epsilon_{p,t}$ is the output of the $p^{\rm{th}}$ relay at time $t$ and the relay gains are $k_1,\ldots,k_p$. 
If we assume that the relay outputs and gains do not change within the considered time period, then it is clear that the rows in $\mathbf{X}$ will be the same and this matrix is therefore singular (since $\rm{det}(\mathbf{X}) = 0$) and cannot be inverted to determine $\mathbf{g}$.

To ensure that the matrix $\mathbf{X}$ is invertible, we propose a solution based on varying the gains of the relays. A simple way to vary the relay gains $\mathbf{K}$ is to introduce a random noise component into the nominal gains for each relay where the noise sequences are independent. For example:
\begin{align}
\label{eq:randomgain}
    \mathbf{K}_t = 2 \mathbf{K}_0 \otimes \mathbf{D}_t
\end{align}
where $\otimes$ is the Kronecker product, $\mathbf{K}_0$ is the nominal gains and $\mathbf{D}_t = [\delta_{1,t},\ldots,\delta_{p,t}]^T$ is a vector of independent uniform random numbers in the range $0$ to $1$. The random variation introduced into the relay gains will then ensure that the estimation of the partial derivatives is possible. Note that the expected value of the stochastic gains is equal to the nominal gains, i.e., $E[\mathbf{K}]=\mathbf{K}_0$.

Randomizing the relay gains allows estimation of $\mathbf{g}$ via Equation~\ref{eq:LS} using a moving window to populate the $\mathbf{X}$ and $\mathbf{Y}$ matrices. The control block sets the relay directions based on the sign of the estimated gradient vector $\mathbf{g}$. Based on this approach, and with cost function minimization, the overall closed loop in Figure~\ref{fig:ss_block} can be represented as the following simplified dynamic equation:
\begin{align}
\label{eq:dynsystem}
    \dot{\boldsymbol{\theta}} = -\mathbf{K} \otimes \sgn(\mathbf{g})
\end{align}
where $\dot{\boldsymbol{\theta}} = [\frac{d\theta_1}{dt},\ldots,\frac{d\theta_p}{dt}]^T$ and $\mathbf{g} = [\frac{\partial y}{\partial\theta_1},\ldots,\frac{\partial y}{\partial\theta_p}]^T$. This characterization is useful in proving convergence and stability as described in the next section.

\subsection{Simple stability proof}
\label{sect:stability}
Assuming that the static map $Q$ is continuously differentiable and its gradient is Lipschitz continuous, Lyaponov's direct method can be used to demonstrate the stability of the characterization presented in Equation~\ref{eq:dynsystem}. First, we choose $V(\boldsymbol{\theta})=Q(\boldsymbol{\theta})-Q(\boldsymbol{\theta}^*)$ where $\boldsymbol{\theta}^*$ is an equilibrium point, which is stable and convergent since $V(\boldsymbol{\theta})>0$ for $\boldsymbol{\theta} \neq \boldsymbol{\theta}^*$ and $V(\boldsymbol{\theta}^*) = 0$ and because its derivative $\frac{d V}{dt}$ is negative (definite or semi-definite):
\begin{align}
\label{eq:lyaponov}
    \frac{dV}{dt} & = \mathbf{g}^T \dot{\boldsymbol{\theta}} \\
                  & = \mathbf{g}^T (-\mathbf{K} \otimes \sgn(\mathbf{g})) \\
                  & = -||(\mathbf{K} \otimes \mathbf{g}||_1
\end{align}
Because the $L_1$ norm is always positive for $\boldsymbol{\theta} \neq \boldsymbol{\theta}^*$ and $\mathbf{K}>0.~\forall k$, then $\frac{dV}{dt}$ is thus always negative. With $\frac{dV}{dt} \leq 0$, $V(\theta)$ and $Q(\theta)$ decrease over time until it reaches a minimum. When $\frac{dV}{dt}=0$, the implication is that $\mathbf{g}=0$, indicating an equilibrium point for $\boldsymbol{\theta}$. 

The algorithm will also yield fast/infinite switching back and forth at the optimal point which, in practical implementations, will be limited due to time delays and physical constraints. In the following section, the method is adapted for dynamic systems, which includes incorporating an intentional hold time on the relays.

With an intentional delay ($T_d$) placed on the switching of the relays, the switching frequency is:
\begin{align}
    f = \frac{1}{2 T_d}
\end{align}
and this leads to an (expected) maximum error $E[\mathbf{e}_{\boldsymbol{\theta}}]$ between $\boldsymbol{\theta}$ and $\boldsymbol{\theta}^*$ when fluctuating around the optimum determined by the nominal gain and the hold time:
\begin{align}
\label{eq:error}
    E[\mathbf{e}_{\boldsymbol{\theta}}] = \mathbf{K}_0 T_d
\end{align}
To summarize, the only configuration requirements for the static map ESC are the nominal gains (in $\bf{K}_0$) for each channel. These parameters can be set based on physical knowledge of the controlled system because they determine the rate of change of each manipulated variable. Systems will usually have physical limits that constrain this rate of change which could thus provide the settings without the need for tuning.

\section{Extension to Dynamic Systems}
\label{sect:dyanmic}
To extend the algorithm beyond a static map to more general dynamic systems, we start by defining a general multi-input-single-output (MISO) nonlinear system as:
\begin{align}
    \dot{x} & = f(x,u) \\
    y & = h(x)
\end{align}
where $x \in \Re^m$ is the state, $u \in \Re^p$ is the input, $y \in \Re$ is the output and $f:\Re^m \times \Re^p \rightarrow \Re^m$ and $h:\Re^m \rightarrow \Re$ are smooth. Assume that there is a control law $u=\alpha(x,\theta)$ parameterized by $\boldsymbol{\theta} \in \Re^p$ and then the closed loop system $\dot{x}=f(x,\alpha(x,\boldsymbol{\theta}))$ has an equilibrium parameterized by $\boldsymbol{\theta}$. 

As per~\cite{krstic2000stability}, we make the following further assumptions:
(1) {\bf differentiable} - there exists a smooth function $l:\Re^p \rightarrow \Re^m$ such that $f(x,\alpha(x,\boldsymbol{\theta}))=0 \iff x=l(\boldsymbol{\theta})$; (2) {\bf stable} - $\forall \theta \in \Re^p$, the equilibrium $x=l(\boldsymbol{\theta})$ of the system is locally exponentially stable uniformly in $\boldsymbol{\theta}$; (3) {\bf convex} - there exists $\boldsymbol{\theta}^* \in \Re^p$ such that $\frac{\partial}{\partial \boldsymbol{\theta}} (h \circ l)(\boldsymbol{\theta}^*) = 0$, and $\frac{\partial^2}{\partial \boldsymbol{\theta}^2} (h \circ l)(\boldsymbol{\theta}^*) = H > 0, H = H^T$. Also, for convenience, we denote $Q(.)=h \circ l(.)$ as the static map.

\begin{figure}[ht]
\centering
\includegraphics[width=3in]{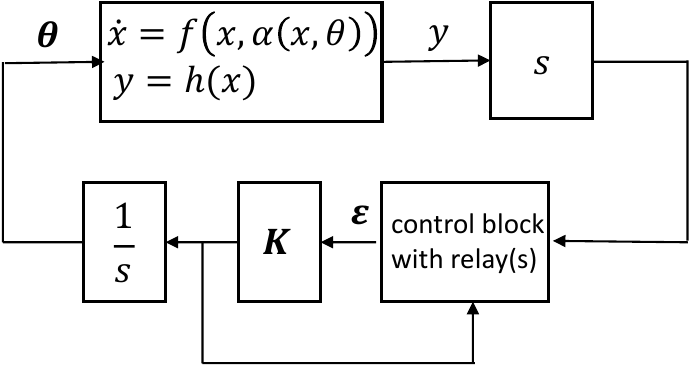}
\caption{Multi-relay ESC for general MISO dynamic system}
\label{fig:dyn_block}
\end{figure}

\begin{figure}[ht]
\centering
\includegraphics[width=3in]{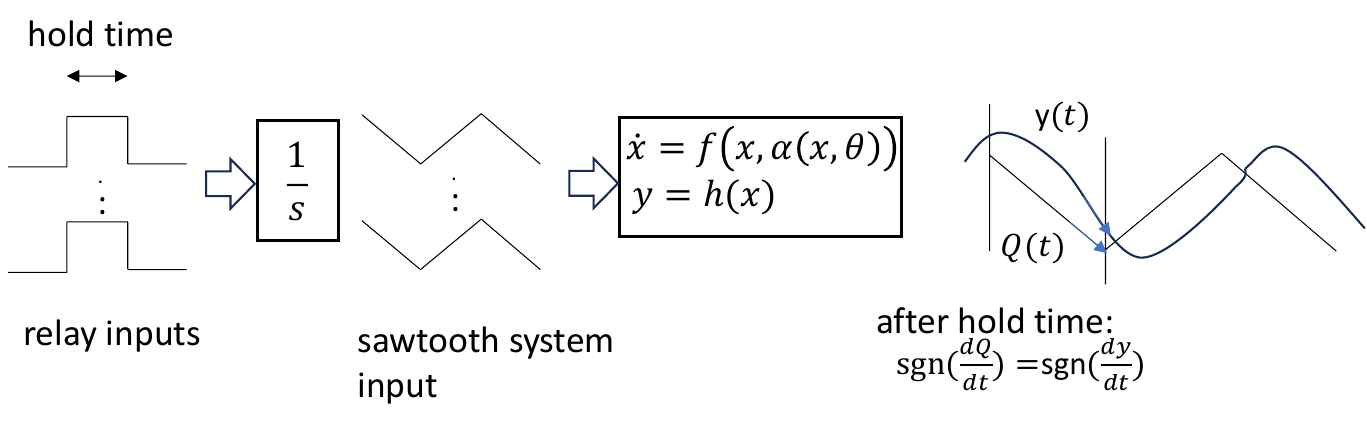}
\caption{Effect of minimum relay hold time}
\label{fig:slopes}
\end{figure}

As is generally true for multivariable ESC, the objective here is to create a strategy that minimizes the cost function (given by the steady state value of $y$) without knowledge of $\boldsymbol{\theta}^*$ or $h$ and $l$. The solution presented in the last section for a static map can be adapted to achieve this goal via a few modifications. First, the relays need to hold their states for long enough so that slope of $y$ has the same sign as its steady-state counterpart, i.e.,  
\begin{align}
    \sgn(\frac{dy}{dt}) = \sgn(\frac{dQ(\boldsymbol{\theta})}{dt})
\end{align}
The effect of holding the relay states for a minimum time is illustrated in Figure~\ref{fig:slopes}. Here the relay outputs pass through the integrator to create a saw-tooth system input. The trajectories of $y(t)$ and its steady state counterpart $Q(t)$ evolve over time in response to the input signal but the assumption is that their signs coincide after the minimum relay hold time. This relay holds time enforces a time scale separation so that the ESC optimizes the system at a time scale slower than the system dynamics.

The system dynamics also introduce auto-correlation into the signals meaning that the implicit assumption in Equation~\ref{eq:estimation} for estimating the new relay directions of having independent samples populating the matrix rows is no longer valid. This problem can be addressed by extending the estimation window so that it includes enough samples over a long enough time to counteract the auto-correlation. This would lead to the $\mathbf{X}$ and $\mathbf{y}$ matrices having more than $p$ rows and would then require a least-squares approach to the parameter estimation to obtain $\mathbf{g}$, i.e.,
\begin{align}
\label{eq:LS}
    \hat{\mathbf{g}} = (\mathbf{X^T X})^{-1} \mathbf{X^T Y}
\end{align}
Extension of the static version of the multi-relay ESC to dynamic systems thus requires the relay(s) hold time to yield more than $p$ rows in $\mathbf{X}$ and $\mathbf{Y}$. The number of samples and time between the samples will necessarily depend on the system dynamics and vis-\`a-vis the time for the system to reach a steady state after a disturbance.

To configure the dynamic version of ESC, the nominal gains need to be set as for the static map, but we also now need the hold time for the relays and the length of time over which to collect samples for the least-squares estimation. Here we propose introducing {\em just one additional parameter} for the dynamic version: $\tau_s$, which is the dominant time constant of the optimized system. 

We propose setting the relay hold time $T_d$ equal to this time constant to ensure time scale separation and quasi-steady-state at switching times. In addition, we propose using recursive-least-squares (RLS) for the estimation in~\ref{eq:LS} with exponential forgetting where the time constant of the forgetting is also set based on $\tau_s$. The algorithm should then be executed in discrete time at a fast enough rate to gather enough samples so that the total number of samples within the period $\tau_s$ is greater than the number of input channels.

\section{Summary of the Algorithm and Configuration Requirements}

As noted earlier, one of the main objectives of this paper is to develop an extremum-seeking control algorithm for MISO systems with minimal configuration requirements making it easier to deploy in practice than alternative perturbation-based approaches. The proposed algorithm requires only a minimal set of parameters for configuration, as shown in Table~\ref{table:config}. These parameters are also easy to determine from the physical properties of the system being optimized.

Note that a single hold time is set for all relays but that the nominal gains $\mathbf{K}_0$ can be set differently depending on the allowable rate of change for each manipulated variable. For the static version, no time-scale separation is required and the algorithm could theoretically be executed as fast as possible with the sample period arbitrarily small. However, the hold time has to equal $p$ (the number of input channels) times the sample period so that enough samples are obtained to enable the solution of Equation~\ref{eq:estimation}. 

\begin{table}[ht]
\caption{MISO ESC configuration parameters}
\label{table:config}
\begin{center}
\begin{tabular}{|l|l|ll|}
\hline
\multirow{2}{*}{\textbf{Parameter}} & \multirow{2}{*}{\textbf{Description}} & \multicolumn{2}{c|}{\textbf{How to set}}                                 \\ \cline{3-4} 
                                    &                                       & \multicolumn{1}{l|}{\textbf{Static}} & \textbf{Dynamic}                  \\ \hline
$T_d$                               & Relays hold time                      & \multicolumn{1}{l|}{$p \Delta t$}             & $\tau_s$                          \\ \hline
$\Delta t$                          & Sample period                         & \multicolumn{1}{l|}{any}   &   $\Delta t \leq \frac{T_d}{p}$                                     \\ \hline
$\mathbf{K}_0$                      & Relay gains                           & \multicolumn{2}{l|}{MV rate of change}                            \\ \hline
$\lambda$                      & RLS forgetting factor                 & \multicolumn{1}{l|}{N/A}             & $\exp({-\frac{\Delta t}{\tau_s}})$ \\ \hline
\end{tabular}
\end{center}
\end{table}

For the dynamic version, the system time constant $\tau_s$\footnote{$\tau_s$ will not be known exactly but overestimating it will only lead to slower convergence compared with potential failure to converge that could result from underestimation due to insufficient time-scale separation.} is used to configure all additional parameters and this can be determined from a step test. The relays are held for a length of time equal to $\tau_s$ for time-scale separation with the sample period small enough so that more than $p$ samples are obtained for estimation of the relay directions from (recursive) least squares (RLS). RLS is used to estimate the gradients with exponentially-weighted forgetting implemented to maintain sensitivity to new data. The forgetting factor for RLS $\lambda$ is calculated also from $\tau_s$ to provide an effective number of samples in the estimation equal or greater than $p$. The full details of the algorithm comprising the control block are shown as Algorithm~\ref{alg:mesc} for the dynamic case.

\begin{algorithm2e}
\caption{Control block algorithm for dynamic systems}
\label{alg:mesc}
\SetAlgoLined
initialize: $\mathbf{P}_1=\gamma \mathbf{I}$ \\
set nominal relay gains: $\mathbf{K}_0$ \\
set initial gradients: $\mathbf{g}_1^T = [0\ldots0]$ \\
set initial relay directions: $\boldsymbol{\epsilon}_1 \in \{-1,1\}$ \\
set counter to zero: $T = 0$ \\
set initial $\mathbf{x}$: $\mathbf{x}_1^T = [\epsilon_{1,1} k_1,\cdots,\epsilon_{p,1} k_p]$ \\
\For{i = $2$ to $N$}
{
increment counter: $T = T + \Delta t$ \\
recursive least-squares estimation:\\
$\mathbf{d}_i = \frac{(\mathbf{P}_{i-1} \mathbf{x}_{i-1})}{\lambda + \mathbf{x^T_{i-1} P_{i-1} x_{i-1}}}$ \\
$\mathbf{P}_i = \frac{1}{\lambda} (\mathbf{P}_{i-1}-\mathbf{d}_i \mathbf{x}^T_{i-1} \mathbf{P}_{i-1})$ \\
$e_i = \frac{dJ}{dt}|_i - \mathbf{x}_{i-1}^T \mathbf{g}_{i-1}$ \\
$\mathbf{g}_i = \mathbf{g}_{i-1} + e_i \mathbf{d}_i$ \\
$\boldsymbol{\epsilon}_{i} = \boldsymbol{\epsilon}_{i-1}$\\
\If{\rm{any}($\sgn(\mathbf{g}) \not= -\sgn(\boldsymbol{\epsilon}_{i-1}))$ AND $T \geq T_{d}$ }
{
reset timer: $T = 0$ \\
set new relay directions: $\boldsymbol{\epsilon}_{i} = -\sgn{(\mathbf{g}_{i})}$\\
}
set relay gains: $\mathbf{K}_i = 2\mathbf{K}_0 \otimes \mathbf{D}_i$ \\
set new $\mathbf{x}$: $\mathbf{x}_i^T = [\epsilon_{1,i} k_{1,i},\cdots,\epsilon_{p,i} k_{p,i}]$ 
}
\end{algorithm2e}

\subsection{Convergence versus persistent excitation}

Note that the convergence rate and oscillation around the optimum are jointly determined by the relay gains. This means that faster convergence can be obtained by increasing the gains but this will also then lead to a greater amplitude of oscillation around the optimal point. This is a disadvantage compared to conventional sinusoidal ESC which allows for these two aspects of performance to be more easily decoupled albeit at the expense of additional configurable parameters. However, a simple solution to this problem is to vary the nominal gains of the relays based on a norm of the estimated gradients. 

Some possible approaches to adjusting the relay gains in this way were explored by the authors in previous work~(\cite{salsbury2020self}). Extending this previous work to the current MISO strategy, we can define a vector $\mathbf{g}_{\rm{abs}} \in \Re^p$ of absolute values of the estimated gradients:
\begin{align}
\mathbf{g}_{\rm{abs}} = [|g_1|,|g_2|,\ldots,|g_p|]  
\end{align}
The adaptive relay gains can then be determined from:
\begin{align}
\label{eq:adaptivegain}
\mathbf{K} = 2 \mathbf{K}_0 \otimes \left(\mathbf{1}_p + \mathbf{g}_{\rm{abs}} + \zeta \mathbf{D} \right)
\end{align}
where $\mathbf{D}$ is a vector of independent uniform random numbers in the range $0$ to $1$ but scaled by a small number $\zeta$. Hence, the relay gains are still subject to some random variation but their magnitude is mainly adjusted according to the magnitude of the estimated gradients. The next section presents some simulation test results that demonstrate the potential of this adaptive gain approach.

\section{Simulation Results}
\label{sect:sim}
Simulation tests were performed for both the static and dynamic versions of the MISO relay-based strategy. For both cases, the static map was of the following quadratic form:
\begin{align}
Q(t) = \frac{1}{2}(\boldsymbol{\theta}(t)-\boldsymbol{\theta}^*)^T \mathbf{H}(\boldsymbol{\theta}(t)-\boldsymbol{\theta}^*)
\end{align}
where $\boldsymbol{\theta}(t)$ is the vector of inputs at time $t$; $\boldsymbol{\theta}^*$ are the optimal values for the inputs; $\mathbf{H}$ is set to the identity matrix. To extend the analysis to the dynamic case, we assumed simple linear dynamics as a Hammerstein formulation  such that:
\begin{align}
\dot{x}(t) & = A x(t) + B Q(t) \\
y(t) & = C x(t)
\end{align}
with $A=-1/\tau_s$, $B=1/\tau_s$, $C=1$. Thus, for the dynamic case, SISO dynamics are applied to the scalar cost function $Q$, and $y$ then becomes the measurement used in the optimization. The dynamic model represents an abstraction of many real-world systems with multiple inputs, a single measurable cost function, and stable dynamics. A simulation was set-up with two input channels $\theta_1$ and $\theta_2$. The optimum values ($\boldsymbol{\theta}^*$) for these two channels were stepped between two different values $[0.2,0.7] \rightarrow [0.8,0.3]$. For the static test, the hold time for the relay was set to 2 samples (equal to the number of input channels) whereas for the dynamic case, the hold time was set equal to the system time constant (10 seconds) with the sample period $\Delta t = 1$ second. Note that the gains for each input channel ($k_1$ and $k_2$) are set to the same values due to the symmetric static map in the simulation model. We tested two gain values for this experiment: $K = 0.01$ and $K = 0.001$.

\begin{figure}[ht]
\centering
\includegraphics[width=0.5\textwidth]{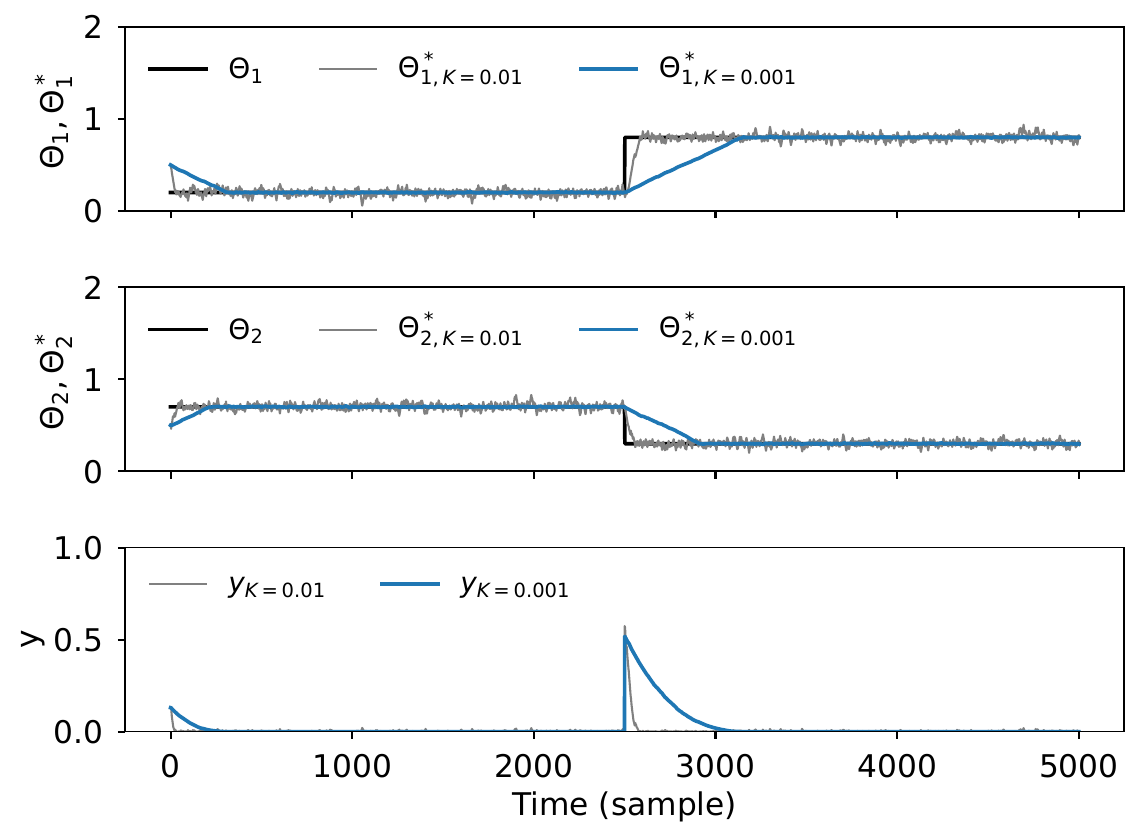}
\caption{Simulation results for static version}
\label{fig:static_response}
\end{figure}

\begin{figure}[ht]
\centering
\includegraphics[width=0.5\textwidth]{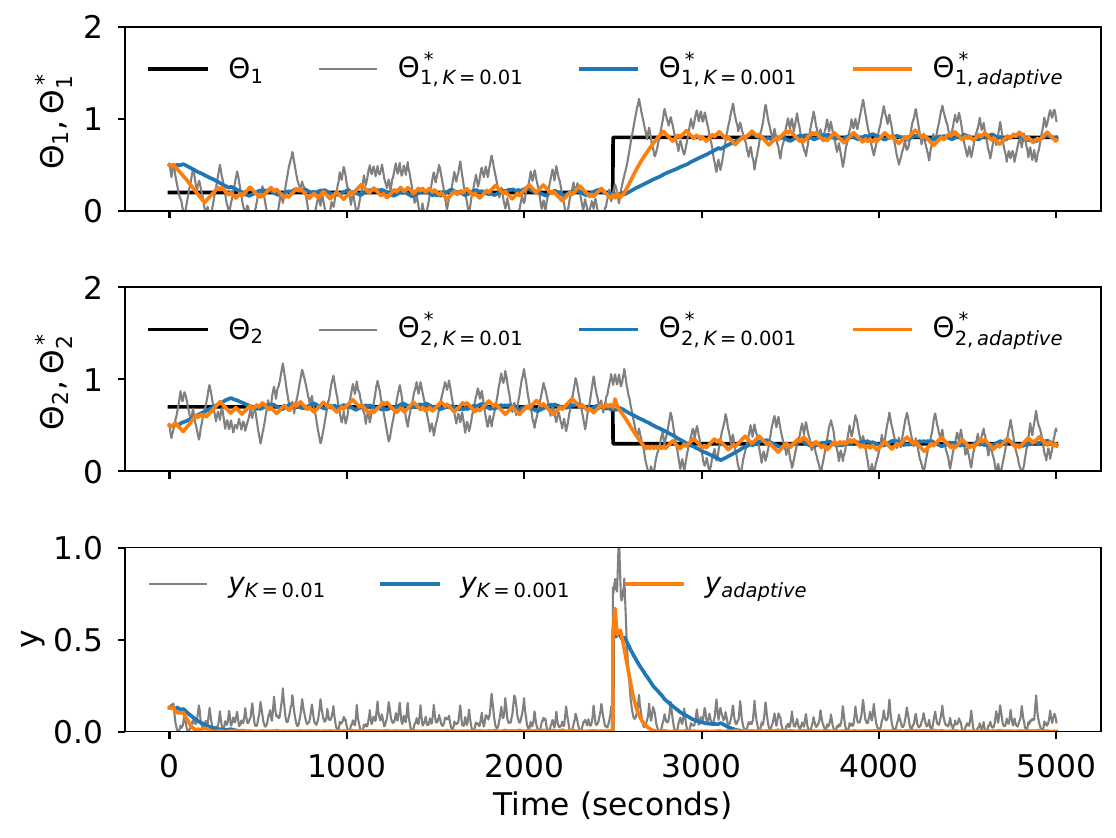}
\caption{Simulation results for dynamic version including adaptive gain approach}
\label{fig:dynamic_response}
\end{figure}

\begin{figure}[ht]
\centering
\vspace{-0.4cm}
\includegraphics[width=0.45\textwidth]{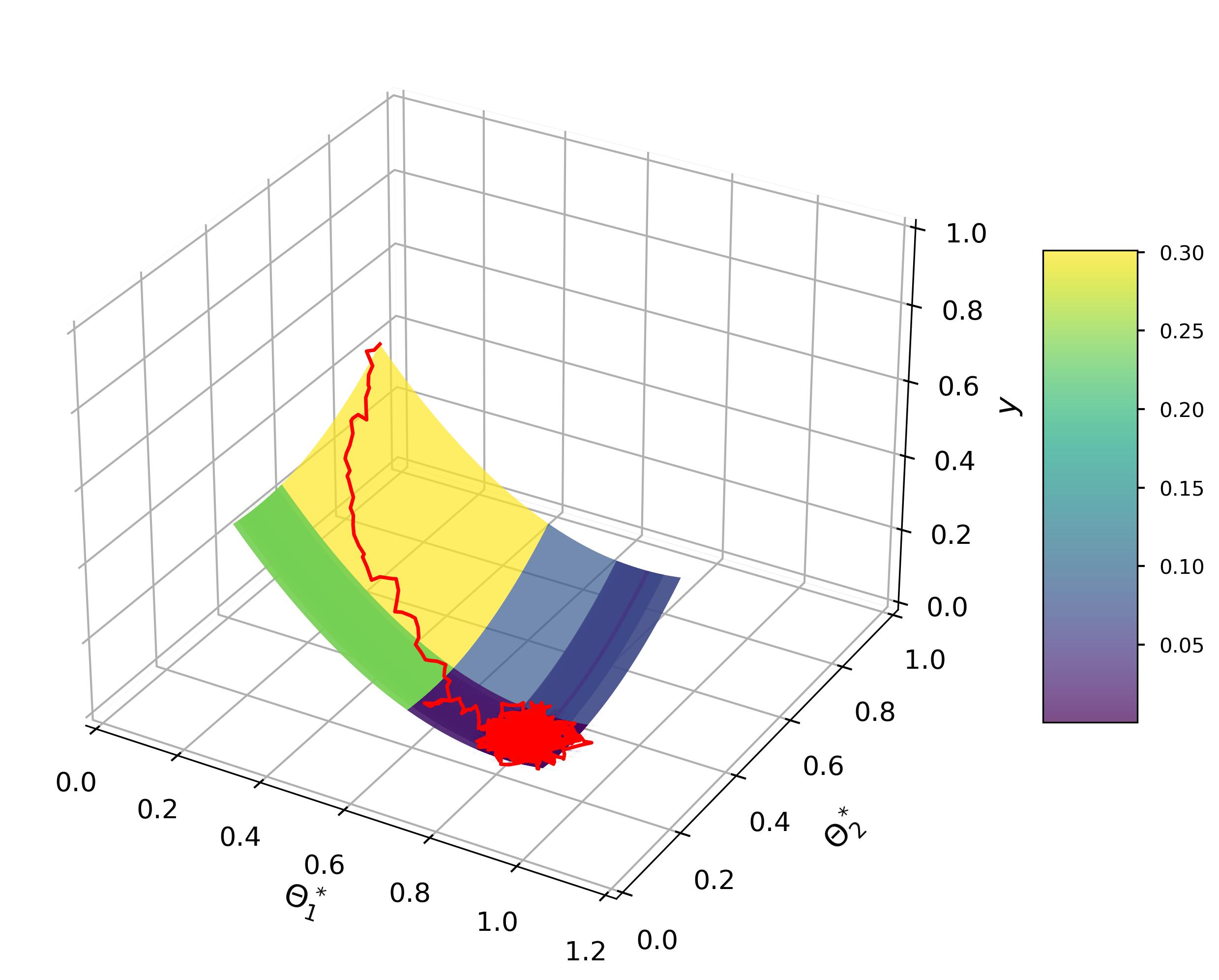}
\caption{Trajectory of objective function $y$ for the static case with $K = 0.01$ towards optimum value}
\label{fig:static_k01trajectory}
\end{figure}

\begin{figure}[ht]
\centering
\vspace{-0.4cm}
\includegraphics[width=0.45\textwidth]{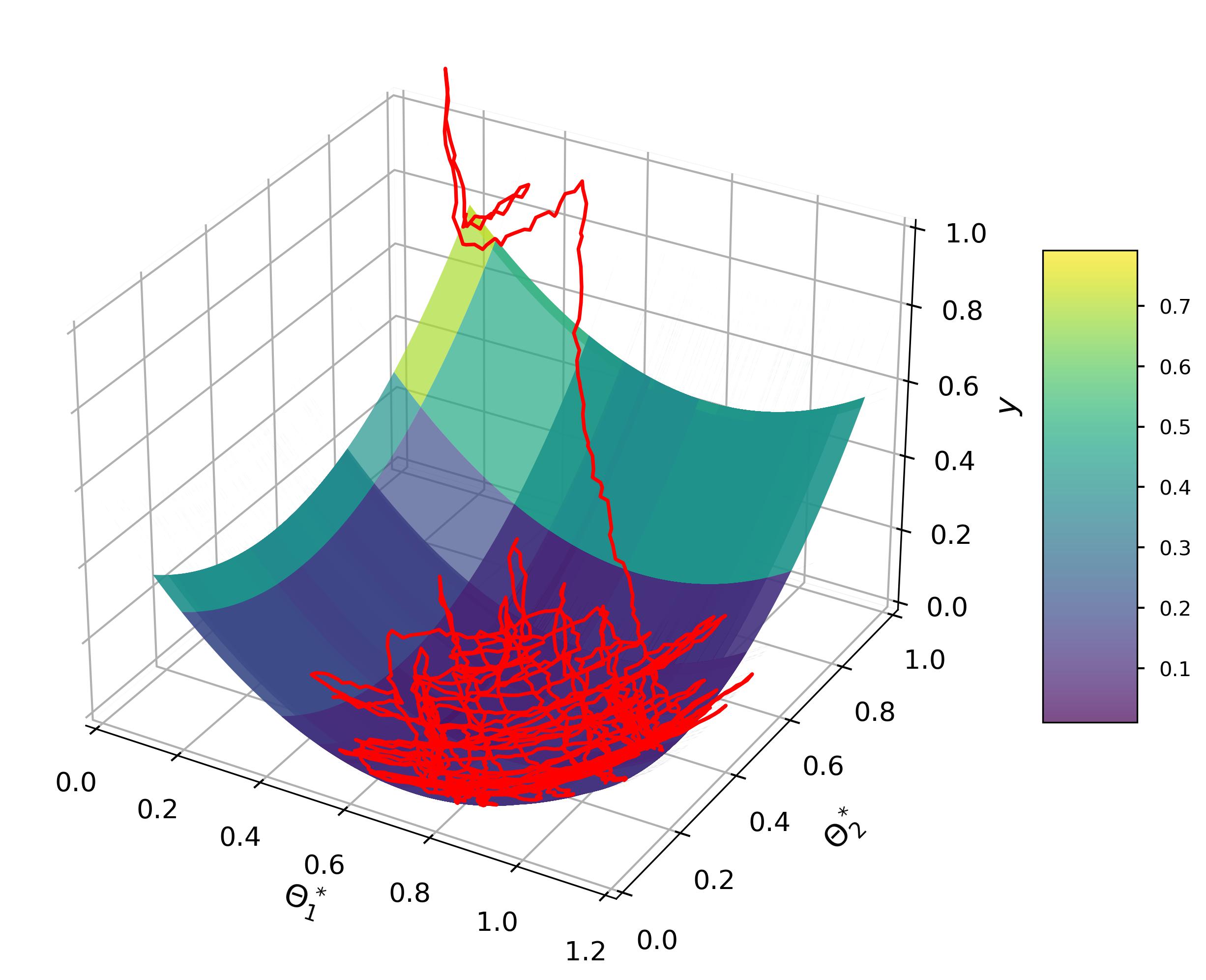}
\caption{Trajectory of objective function $y$ for the dynamic case with $K = 0.01$ towards optimum value}
\label{fig:dynamic_k01trajectory}
\end{figure}

\begin{figure}[ht]
\centering
\vspace{-0.4cm}
\includegraphics[width=0.45\textwidth]{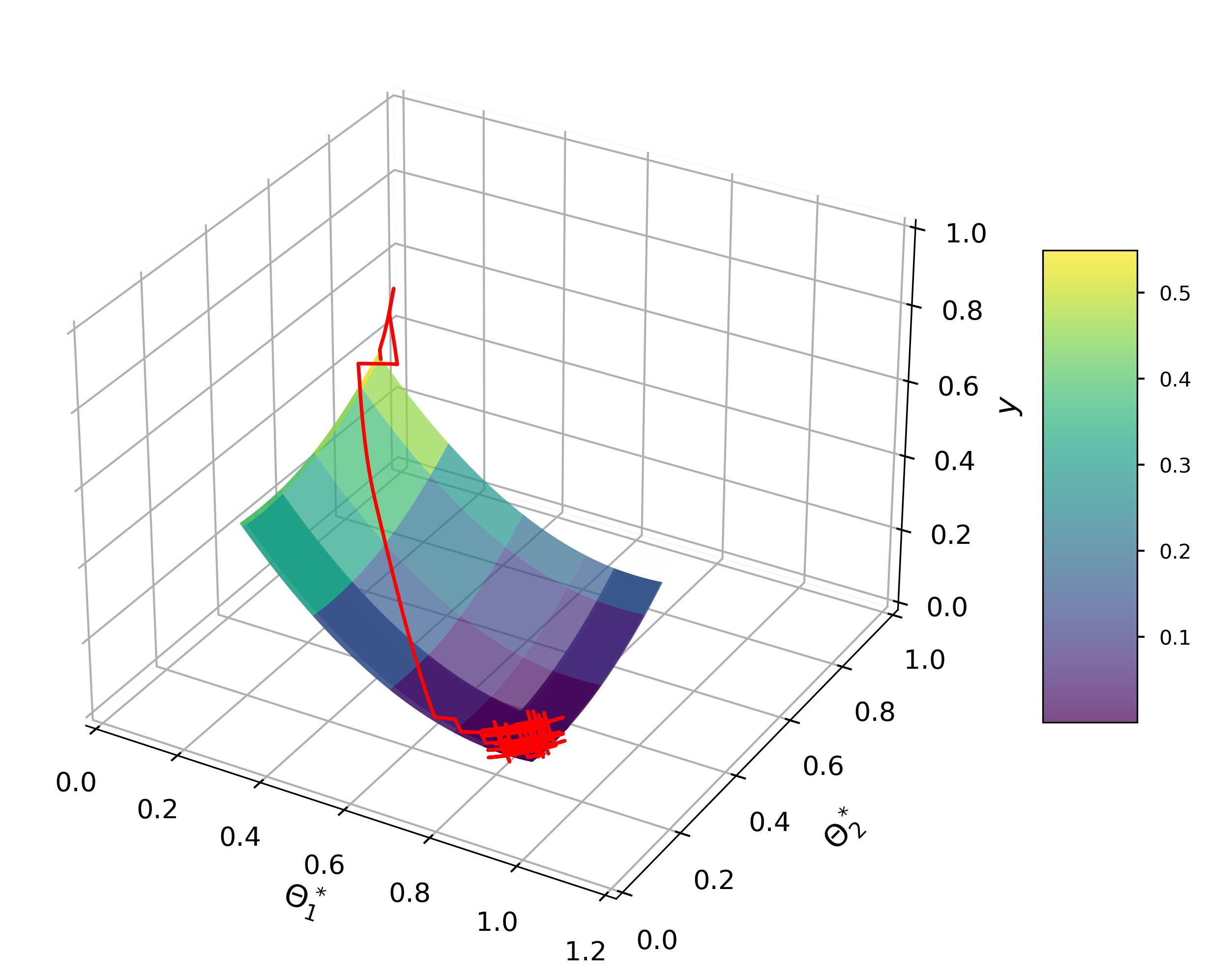}
\caption{Trajectory of objective function $y$ for the dynamic case with the adaptive gain approach towards optimum value}
\label{fig:adaptive_trajectory}
\end{figure}

Test results for both static and dynamic cases are shown in Figure~\ref{fig:static_response} and Figure~\ref{fig:dynamic_response} where the two upper subplots show the inputs ($\boldsymbol{\theta_{K=0.01}}$, $\boldsymbol{\theta_{K=0.001}}$) in relation to the optimal values ($\boldsymbol{\theta^*}$). The bottom subplot shows the cost function $y$. When $K$ is smaller, such as $K=0.001$, it takes longer to reach the target values, but the oscillation around the target value is smaller. Conversely, when $K=0.01$, the larger $K$ leads to faster movement toward the target value, at the expense of larger oscillations around it. Thus, there is a trade-off between these features when choosing the value of $K$. Also, the dynamic case experiences a larger oscillation around the optimum for equivalent relay gains because of the longer hold time in terms of the number of samples, as expected from Equation~\ref{eq:error}. 

Figure~\ref{fig:dynamic_response} also shows one test result using the adaptive gain approach in Equation~\ref{eq:adaptivegain}. For this test, $\zeta$ was set to $0.001$ in Equation~\ref{eq:adaptivegain}. The adaptive gain approach decouples the speed of convergence from the amplitude of oscillation by allowing the relay gains to be higher during convergence and lower when oscillating. The test result demonstrates its enhanced performance by effectively addressing the trade-off, achieving faster convergence and correspondingly less oscillation than the fixed gain cases.

The optimization maps are shown in Figure~\ref{fig:static_k01trajectory} -- Figure~\ref{fig:adaptive_trajectory} using color variations to represent the search regions with the input values. They also show the trajectory of the input after stepping $\boldsymbol{\theta}$, where the red line is the path from the initial conditions to the optimum value. Initially, $\boldsymbol{\theta^*_{1}}$ and $\boldsymbol{\theta^*_{2}}$ start near 0.2 and 0.7, then approach the target values 0.8 and 0.3. The static version in Figure~\ref{fig:static_k01trajectory} has a smooth trajectory toward the optimum and small variation at convergence. In contrast, the dynamic version in Figure~\ref{fig:dynamic_k01trajectory} follows a noisier path toward and around the optimum. Adding the adaptive gain improves overall performance as shown in Figure~\ref{fig:adaptive_trajectory} with much less variation at the optimum while still converging quickly.

\section{Conclusion}
\label{sect:conclusions}
The SISO relay-based version of extremum-seeking control is attractive for a number of reasons including ease of configuration, no external dither signal, and intuitive operation. This paper has extended the idea to MISO systems with the primary goal of creating a new multivariable extremum-seeking control solution that is easier to deploy in practice than alternative algorithms that have been reported in the literature. The new algorithm was developed based on the multivariable chain rule and stochastic relay gains to determine the direction of the relays via a least squares estimation. The approach was first developed for a static system and then extended to more general dynamic systems with the latter including time scale separation.

Simulation was used to test both the static and dynamic versions of the algorithm and the results demonstrated a good trade-off between the level of persistent excitation and convergence speed. The basic version of the proposed algorithm has fixed nominal gains for the relays which means that convergence speed and the amplitude of persistent oscillation at the optimum are coupled. Increasing gains can thus increase convergence speed but then lead to worse oscillations. This is a disadvantage of the proposed approach but it was also shown how this could be remedied by allowing the nominal gains to vary as a function of the norm of the estimated cost function gradients. One simulation result was presented to demonstrate the potential of this adaptive gain approach but further work is needed to investigate this approach.

%\begin{ack}   
%\end{ack}

\bibliography{refs}     

@article{salsbury2020self,
  title={Self-perturbing extremum-seeking controller with adaptive gain},
  author={Salsbury, Timothy I and House, John M and Alcala, Carlos F},
  journal={Control Engineering Practice},
  volume={101},
  pages={104456},
  year={2020}
}

@inproceedings{olalla2007analysis,
  title={Analysis and comparison of extremum seeking control techniques},
  author={Olalla, Carlos and Arteaga, Maria Isabel and Leyva, Ramon and El Aroudi, Abdelali},
  booktitle={2007 IEEE International Symposium on Industrial Electronics},
  pages={72--76},
  year={2007},
  organization={IEEE}
}

@inproceedings{blackman1962extremum,
  title={Extremum-seeking regulators},
  author={Blackman, PF},
  booktitle={An exposition of adaptive control},
  year={1962},
  organization={Macmillan}
}

@inproceedings{tan2010extremum,
  title={Extremum seeking from 1922 to 2010},
  author={Tan, Ying and Moase, William H and Manzie, Chris and Ne{\v{s}}i{\'c}, Dragan and Mareels, Iven MY},
  booktitle={Proceedings of the 29th Chinese control conference},
  pages={14--26},
  year={2010},
  organization={IEEE}
}

@article{leyva2006mppt,
  title={MPPT of photovoltaic systems using extremum-seeking control},
  author={Leyva, Ramon and Alonso, Corinne and Queinnec, Isabelle and Cid-Pastor, Angel and Lagrange, Denis and Martinez-Salamero, Luis},
  journal={IEEE transactions on aerospace and electronic systems},
  volume={42},
  number={1},
  pages={249--258},
  year={2006}
}

@inproceedings{ariyur2002analysis,
  title={Analysis and design of multivariable extremum seeking},
  author={Ariyur, Kartik B and Krstic, Miroslav},
  booktitle={Proceedings of the 2002 American Control Conference},
  volume={4},
  pages={2903--2908},
  year={2002},
  organization={IEEE}
}

@article{ghaffari2012multivariable,
  title={Multivariable Newton-based extremum seeking},
  author={Ghaffari, Azad and Krsti{\'c}, Miroslav and Ne{\v{s}}i{\'c}, Dragan},
  journal={Automatica},
  volume={48},
  number={8},
  pages={1759--1767},
  year={2012}
}

@article{dong2015self,
  title={Self-optimizing control of air-source heat pump with multivariable extremum seeking},
  author={Dong, Liujia and Li, Yaoyu and Mu, Baojie and Xiao, Yan},
  journal={Applied Thermal Engineering},
  volume={84},
  pages={180--195},
  year={2015}
}

@article{krstic2000stability,
  title={Stability of extremum seeking feedback for general nonlinear dynamic systems},
  author={Krstic, Miroslav and Wang, Hsin-Hsiung},
  journal={Automatica},
  volume={36},
  number={4},
  pages={595--602},
  year={2000}
}

@article{aminde2021multivariable,
  title={Multivariable extremum seeking control via cyclic search and monitoring function},
  author={Aminde, Nerito Oliveira and Oliveira, Tiago Roux and Hsu, Liu},
  journal={International Journal of Adaptive Control and Signal Processing},
  volume={35},
  number={7},
  pages={1217--1232},
  year={2021}
}

@article{korovin1974using,
  title={Using sliding modes in static optimization and nonlinear programming},
  author={Korovin, SK and Utkin, VI},
  journal={Automatica},
  volume={10},
  number={5},
  pages={525--532},
  year={1974}
}

@article{oliveira2011output,
  title={Output-feedback global tracking for unknown control direction plants with application to extremum-seeking control},
  author={Oliveira, Tiago Roux and Hsu, Liu and Peixoto, Alessandro Jacoud},
  journal={Automatica},
  volume={47},
  number={9},
  pages={2029--2038},
  year={2011}
}

@article{oliveira2012global,
  title={Global real-time optimization by output-feedback extremum-seeking control with sliding modes},
  author={Oliveira, Tiago Roux and Peixoto, Alessandro Jacoud and Hsu, Liu},
  journal={Journal of the Franklin Institute},
  volume={349},
  number={4},
  pages={1397--1415},
  year={2012}
}

@article{peixoto2020multivariable,
  title={Multivariable extremum-seeking by periodic switching functions with application to raman optical amplifiers},
  author={Peixoto, Alessandro Jacoud and Oliveira, Tiago Roux and Pereira-Dias, Diego and Monteiro, Jo{\~a}o Carlos},
  journal={Control Engineering Practice},
  volume={96},
  pages={104278},
  year={2020}
}

@article{oliveira2007control,
  title={Control of uncertain nonlinear systems with arbitrary relative degree and unknown control direction using sliding modes},
  author={Oliveira, TR and Peixoto, AJ and Nunes, EVL and Hsu, L},
  journal={International Journal of Adaptive Control and Signal Processing},
  volume={21},
  number={8-9},
  pages={692--707},
  year={2007}
}

@article{scheinker2024100,
  title={100 years of extremum seeking: A survey},
  author={Scheinker, Alexander},
  journal={Automatica},
  volume={161},
  pages={111481},
  year={2024}
}
                                                             
\end{document}